\documentclass{article}
\usepackage{spconf,amsmath,graphicx,hyperref}
\usepackage{multirow}
\usepackage{cite}
\usepackage{subcaption}

\title{Time-frequency localization of bird calls in dense soundscapes}

\name{Simen Hexeberg$^{2}$, Fanghui Tong$^{3}$, Hari Vishnu$^{1}$, Mandar Chitre$^{1,2}$}
\address{
$^{1}$Tropical Marine Science Institute, National University of Singapore\\
$^{2}$Dept. of Electrical and Computer Engineering, National University of Singapore\\
$^{3}$School of Computing, National University of Singapore
}

\begin{document}

\maketitle

\begin{abstract}
Passive acoustic monitoring enables large-scale wildlife observation. Most bioacoustic classifiers predict species presence in a time window without localizing vocalizations precisely in time or frequency, limiting downstream analyses. We formulate time-frequency localization of bird calls as object detection on spectrograms and compare three computer vision model families (YOLO11, SAM 3, RF-DETR) against a non-learnable baseline. We introduce Intersection over Minimum (IoMin), an evaluation metric that better handles ambiguous acoustic boundaries than IoU. We also open-source a browser-based tool for efficient bounding-box labeling. The best RF-DETR model nearly doubles baseline performance on in-distribution, dense soundscapes from Singapore (83.4\% vs. 42.1\% IoMin@50 F1-score) and generalizes better to out-of-distribution recordings from Hawaii (63.2\% vs. 48.6\%). These results indicate that fine-tuned computer-vision models are well suited for time-frequency localization of bird vocalizations in complex soundscapes.
\end{abstract}



\section{Introduction}
\label{sec:intro}

Combining Passive Acoustic Monitoring (PAM) with machine-learning classifiers enables biodiversity monitoring at spatial and temporal scales infeasible with traditional manual surveys. Most bioacoustic classifiers operate on fixed-duration broadband spectrograms~\cite{schwinger2025}. In dense soundscapes, such \textit{global-context} inputs may contain simultaneous vocalizations from multiple species and substantial background noise, making classification difficult and increasing training data requirements. Moreover, global-context classifiers do not localize predictions in time and frequency, limiting interpretability and downstream analyses. 

In contrast, \textit{local-context} approaches enable measurements of call characteristics such as duration, bandwidth, and amplitude, facilitating studies of how animals modify vocal behavior in response to changes in their acoustic environment~\cite{both2012, hopkins2023, farina2012}. We previously proposed a local-context bird classification method, where time-frequency events (TFEs) are extracted using an energy-based detector and subsequently embedded and classified~\cite{hexeberg2025}. This approach performed well across many species with little labeled data. However, the pipeline depends heavily on the first-stage TFE detector: false positives increase the burden on the downstream classifier, while false negatives (missed events) cannot be recovered by later stages.

\begin{figure*}[htbp]
\centering
\begin{subfigure}[t]{0.48\textwidth}
    \centering
    \includegraphics[width=\linewidth]{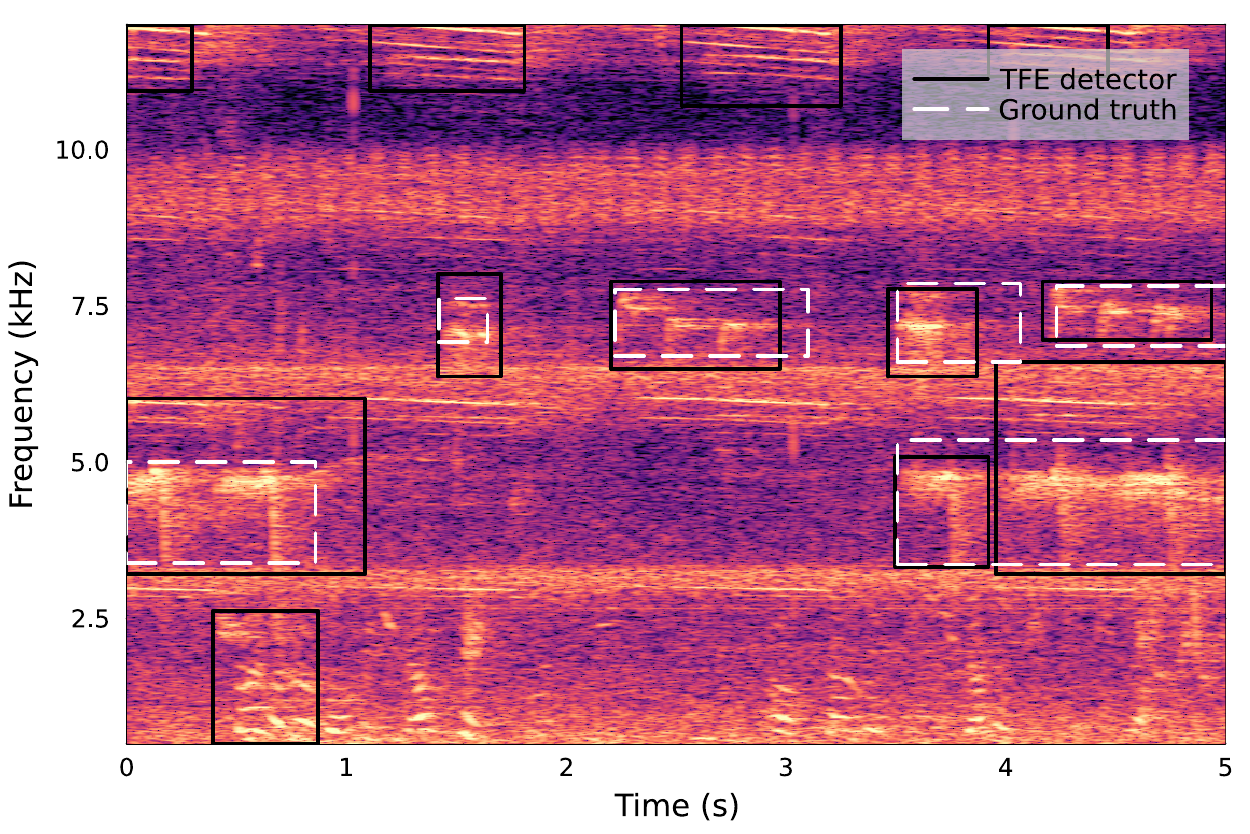}
\end{subfigure}\hspace{0.01\textwidth}
\begin{subfigure}[t]{0.48\textwidth}
    \centering
    \includegraphics[width=\linewidth]{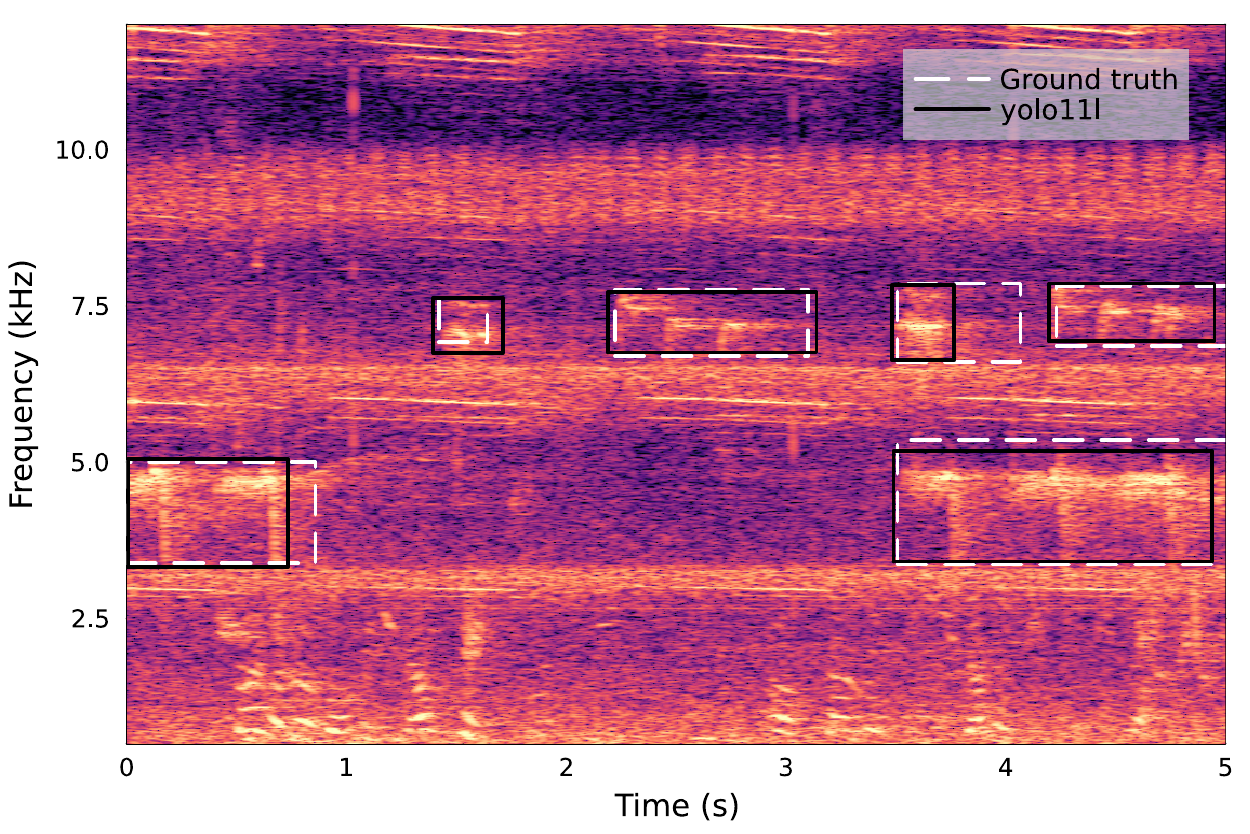}
\end{subfigure}
\caption{TFE detector (left) and YOLO11l (right) predictions on the same Singapore recording. The TFE detector detects non-bird sounds (insect noise, human speech) and produces imprecise boundaries, both of which YOLO avoids.}
\label{fig:tfe-fail-sg}
\end{figure*}

Non-learnable, energy-based TFE detectors~\cite{Brandes2008, Abbasi2025, hexeberg2025} are attractive because they require no training. However, they extract narrowband frequency-modulated (FM) signals and cannot distinguish bird vocalizations from other FM signals. For instance, in Singapore's complex soundscape, the TFE detector in~\cite{hexeberg2025} detects bird calls but also insect noise and human speech (Figure~\ref{fig:tfe-fail-sg}). Moreover, the event boundaries are not robust to overlapping signals. This motivates learnable methods that better localize bird calls in time and frequency amid ambient noise and FM signals from non-target sources.

We formulate TFE extraction as a two-dimensional time-frequency localization problem and evaluate three computer vision model families: YOLO11~\cite{yolov11}, Segment Anything Model 3 (SAM 3)~\cite{sam3}, and RF-DETR~\cite{rfdetr}. YOLO has previously been applied to bioacoustic sound event detection~\cite{Hexeberg2023, Escobar2023, Huang2025}, but its application to bird vocalizations remains limited. The SAM family has seen even less use in bioacoustics, but the initial version was recently applied to dolphin whistle contour detection~\cite{Zhang2025}. RF-DETR, to the best of our knowledge, has not previously been used for sound event detection. We compare all three model families against the energy-based detector from~\cite{hexeberg2025} on bird vocalizations recorded in Singapore and Hawaii.

This work makes three contributions: (1) benchmarking an energy-based TFE detector against three object detection model families for localizing bird vocalizations in dense soundscapes; (2) introducing Intersection over Minimum (IoMin), an IoU adaptation that accounts for ambiguity in acoustic event boundaries; and (3) releasing a browser-based annotation tool for acoustic-event labeling\footnote{https://org-arl.github.io/birdwatch-public/}, along with source code and model checkpoints for reproducibility\footnote{https://github.com/org-arl/birdwatch-public}.

\section{Methodology}

\subsection{Spectrogram Generation}
\label{subsec:specgram}
YOLO11, SAM 3, and RF-DETR are pretrained on natural images. To apply them to audio, we convert waveforms to time-frequency representations via the short-time Fourier transform (STFT). Recordings are split into 6-second windows with 1-second overlap, and the frequency range is restricted to $0.5$--$12~\mathrm{kHz}$ to capture most bird vocalizations while excluding out-of-band noise. For a $44.1~\mathrm{kHz}$ sampling rate, we use $N_{\mathrm{FFT}} = 4096$, a Hann window, and a hop length of 259. This yields near-square spectrograms. With minor downsampling and zero-padding, we reach a target size of $1024\times1024$ for YOLO11 and RF-DETR, and $1008\times1008$ for SAM 3.

To improve visibility of low-SNR vocalizations, STFT magnitudes are converted to log-power, clipped to the $[1,\,99.8]$ percentile range, and gamma-compressed ($\gamma=0.85$). Since the models require three-channel inputs, the single-channel spectrograms are converted to RGB using the magma colormap, which performed better than replicating grayscale across channels for YOLO, possibly because its color statistics better match those of the COCO pretraining dataset~\cite{coco}.

\subsection{Datasets and Annotation}
\label{subsec:data}
We use two datasets (Table~\ref{tab:dataset-comparison}): one recorded in Singapore for training and in-distribution (ID) evaluation, and one recorded in Hawaii for out-of-distribution (OOD) evaluation of generalization to new species and soundscapes.

\begin{table}[ht]
\centering
\small
\renewcommand{\arraystretch}{1.1}
\caption{Comparison of the in-distribution (ID) and out-of-distribution (OOD) datasets.}
\label{tab:dataset-comparison}
\begin{tabular}{l|cc}
\hline
\textbf{Dataset} & \textbf{Singapore} & \textbf{Hawaii} \\
\hline\hline
Role                      & ID & OOD \\
Usage                     & train/val/test & test \\
Recording sites           & 2 & 4 \\
Annotation count          & 18\,095 & 81\,691 \\
Audio duration            & 4.5~hours & 51~hours \\
Annotation density (annotations/min) & 68 & 27 \\
\hline
\end{tabular}
\end{table}

To annotate and analyze recordings efficiently, we developed and open-sourced an annotation tool called BirdWatch. It is designed for fine-tuning bounding-box detectors such as YOLO on acoustic data. It runs in the browser and supports bounding-box annotation and performance evaluation. Playback can also be restricted to a box's time-frequency extent, allowing annotators to listen to isolated calls in noisy soundscapes. We convert 4.5~hours of recordings from two sites in the Singapore Botanic Gardens~\cite{hexeberg2025} into 6-second spectrograms with 1-second overlaps and manually annotate them, yielding 18{,}095 labels. Consecutive blocks of ten spectrograms are split into training, validation, and test sets in a 7:1:2 ratio. The 1-second overlaps between data splits are masked to prevent data leakage. 

The Hawaii dataset~\cite{hawaii-data} is used only for testing and was selected for its ecological and acoustic differences from Singapore: whereas the Singapore Botanic Gardens is a low-elevation equatorial site in a densely populated urban environment, the Hawaiian sites are remote, high-elevation locations with different habitats, avifauna, and soundscape. To align with Singapore, the 27 labeled species are treated as a single ``bird present" class, and annotations are cropped to 6-second, $0.5$--$12$~kHz spectrograms with one second overlap.

\subsection{Models}

\textbf{YOLO.} We evaluate all five YOLO11 sizes---nano (n), small (s), medium (m), large (l), and extra-large (x). Due to space constraints, we report only two: YOLO11n for the best accuracy-compute trade-off and YOLO11l for the best overall performance. The models are initialized from COCO-pretrained weights and fine-tuned on the Singapore training set using YOLO's default learning rate ($1 \times 10^{-2}$), weight decay ($5 \times 10^{-4}$), and Non-Maximum Suppression (NMS) IoU threshold (0.7). NMS reduces redundant predictions by suppressing boxes whose IoU with a higher-confidence box exceeds the threshold. We use YOLO's default augmentation pipeline, as a preliminary comparison on the Singapore dataset showed no benefit from restricting augmentations to acoustically motivated transformations.\\

\noindent  \textbf{Segment Anything Model.} SAM 3 is a large pretrained foundation model for text-conditioned object detection and segmentation. Unlike YOLO11, which is predominantly a convolutional architecture, SAM 3 employs transformer-based vision and text encoders and a transformer-based detection module. With 848M parameters, it is also substantially larger, with 326 times as many parameters as YOLO11n. We use SAM 3 for bounding-box detection (not segmentation), and follow its official fine-tuning configuration. However, we freeze the vision and text encoders and fine-tune only the detection module (30.4M parameters). We train with a batch size of 8, learning rate of $2 \times 10^{-4}$, and early stopping based on validation AP. At inference, NMS with an IoU threshold of 0.7 is applied for consistency with YOLO11. 

For our single-class task, we use a fixed ``bird vocalization'' prompt. As SAM 3 is not pretrained on spectrograms, the resulting text embedding is unlikely to aid detection. To verify this, we replaced the prompt with a random string at inference and observed only about a 2-percentage-point drop in IoMin@50 F1-score, suggesting that the model largely ignores the text embedding.\\

\noindent \textbf{RF-DETR.} RF-DETR is a transformer-based object detector. During pretraining, it performed an architecture search over thousands of configurations in a single training run, rather than training each configuration separately. Different architectures can then be selected at inference. The authors provide six default architectures that span different model sizes and accuracy-latency trade-offs. We fine-tune and evaluate the Nano and Large architectures with default settings and no additional hyperparameter tuning.\\

\noindent \textbf{Baseline Model.} We compare all models against the non-learnable, energy-based TFE detector from~\cite{hexeberg2025}. Rather than learning from labeled examples, the TFE detector extracts candidate regions directly from spectrograms using a fixed three-stage pipeline. First, the spectrogram is normalized independently at each frequency bin using the interquartile range as a robust local noise-floor estimate, highlighting energy elevated relative to its frequency band regardless of global SNR. Second, watershed segmentation separates the spectrogram into connected high-energy regions. Third, regions with time-frequency shapes uncharacteristic of bird vocalizations are filtered using heuristic rules.

\subsection{Evaluation Metrics}
\label{subsec:metrics}
Unlike natural objects in images, vocalization boundaries in spectrograms are inherently ambiguous: calls may fade gradually in time and frequency, and call sequences may be annotated as one or multiple events. For example, the bird vocalizations in Figure~\ref{fig:yolo-pred-hawaii-1} are annotated as four events, whereas YOLO11l produces seven predictions. Neither is incorrect, but under IoU@50, only the leftmost prediction is a true positive (TP). The remaining six predictions have IoU below 50\% and are considered false positives (FPs), while the three unmatched ground-truths become false negatives. We therefore make two acoustically motivated modifications to standard TP and FP definitions. First, we complement IoU with IoMin, defined as the ratio of the intersection area, $A_\cap$, to the smaller of the predicted and ground-truth box areas, $A_p$ and $A_{gt}$:
\begin{equation}
    \text{IoMin} = \frac{A_\cap}{\min(A_p,\ A_{gt})}
    \label{eq:iomin}
\end{equation}
IoMin does not penalize partial capture of a target, or predictions exceeding it. To prevent the models from exploiting this by predicting arbitrarily large boxes, IoMin is used only for evaluation, not training. Second, we ignore duplicate TPs: when multiple predictions match the same ground-truth above the overlap threshold, only one counts as a TP and the rest are ignored rather than penalized as FPs. Together, these modifications yield four TPs and zero FPs under IoMin@50 for the above example, better reflecting the underlying detection performance.

Since the optimal confidence threshold is model- and application-dependent, we report mean Average Precision (mAP) under IoU and IoMin, along with maximum F1-score and corresponding precision and recall. As the TFE detector produces no confidence scores, we report only precision, recall, and F1-score.


\section{Results}

\begin{figure}[htbp]
\centerline{\includegraphics[width=\columnwidth]{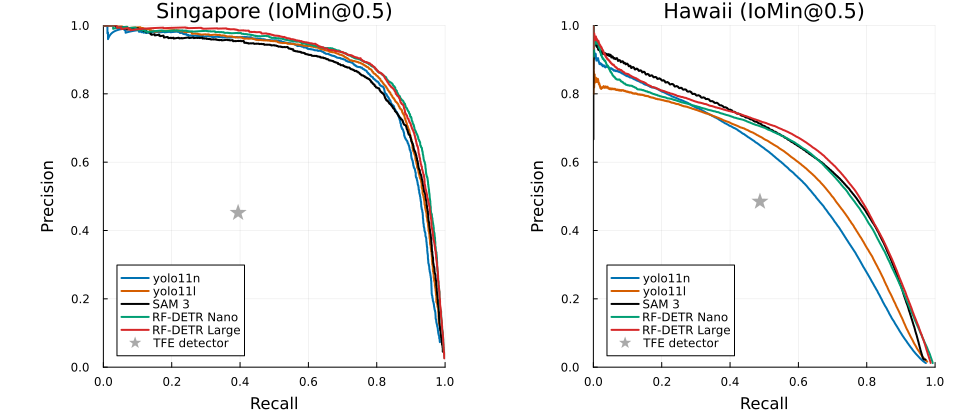}}
\caption{Precision-recall curves for the best-performing training runs, selected by validation F1-score. The gray star indicates baseline performance.}
\label{fig:pr-curve}
\end{figure}

\begin{table*}[htbp]
\centering
\renewcommand{\arraystretch}{1.3}
\footnotesize
\resizebox{\textwidth}{!}{%
\begin{tabular}{l|l|c|c|cccc|cccc}
\hline
\textbf{Dataset} & \textbf{Method} & \textbf{Model Params} & \textbf{Training runs}
  & \multicolumn{4}{c|}{\textbf{IoU@50}}
  & \multicolumn{4}{c}{\textbf{IoMin@50}} \\
\cline{5-12}
 & & & 
  & \textbf{mAP} & \textbf{F1} & \textbf{Prec.} & \textbf{Recall}
  & \textbf{mAP} & \textbf{F1} & \textbf{Prec.} & \textbf{Recall} \\
\hline\hline
\multirow{6}{*}{\shortstack[l]{Singapore\\(ID)}}
 & TFE detector~\cite{hexeberg2025}     & N/A     & N/A
   & N/A              & $14.9$           & $15.5$           & $14.4$
   & N/A              & $42.1$           & $45.2$           & $39.4$           \\
 & YOLO11n & 2.6M & 5
   & $67.3 \pm 0.4$   & $65.8 \pm 0.3$   & $70.4 \pm 2.3$   & $61.9 \pm 1.4$
   & $87.2 \pm 0.3$   & $81.7 \pm 0.2$   & $83.2 \pm 1.8$   & $80.3 \pm 1.5$   \\
 & YOLO11l & 25.3M & 5
   & $67.4 \pm 0.9$   & $66.5 \pm 0.7$   & $69.5 \pm 1.5$   & $63.8 \pm 1.1$
   & $87.5 \pm 0.5$   & $81.8 \pm 0.7$   & $82.1 \pm 1.3$   & $81.6 \pm 1.0$   \\
 & RF-DETR Nano & 30.5M & 5
   & $71.5 \pm 0.6$   & $69.4 \pm 0.3$   & $\mathbf{73.8 \pm 1.2}$   & $65.6 \pm 1.1$
   & $89.9 \pm 0.2$   & $\mathbf{83.4 \pm 0.3}$   & $84.1 \pm 1.4$   & $\mathbf{82.8 \pm 1.7}$   \\
 & RF-DETR Large & 33.9M & 5
   & $\mathbf{71.9 \pm 0.3}$   & $\mathbf{69.6 \pm 0.7}$   & $73.0 \pm 2.0$   & $\mathbf{66.5 \pm 1.5}$
   & $\mathbf{90.0 \pm 0.3}$   & $\mathbf{83.4 \pm 0.2}$   & $\mathbf{84.4 \pm 1.5}$   & $82.6 \pm 1.7$   \\
 & SAM 3 & 848M & 5
   & $64.2 \pm 0.6$   & $64.4 \pm 0.7$   & $67.1 \pm 1.6$   & $62.0 \pm 0.8$
   & $86.5 \pm 0.5$   & $80.6 \pm 0.5$   & $80.7 \pm 1.0$   & $80.5 \pm 0.5$   \\
\hline
\multirow{6}{*}{\shortstack[l]{Hawaii\\(OOD)}}
 & TFE detector~\cite{hexeberg2025}     & N/A     & N/A
   & N/A              & $10.3$           & $8.7$            & $12.5$
   & N/A              & $48.6$           & $48.5$           & $48.7$           \\
 & YOLO11n & 2.6M & 5
   & $6.6 \pm 1.0$    & $16.4 \pm 1.1$   & $15.6 \pm 1.3$   & $17.2 \pm 1.1$
   & $53.2 \pm 2.2$   & $55.9 \pm 1.5$   & $56.2 \pm 1.1$   & $55.7 \pm 2.8$   \\
 & YOLO11l & 25.3M & 5
   & $9.0 \pm 0.6$    & $19.1 \pm 0.7$   & $18.8 \pm 1.3$   & $19.4 \pm 0.6$
   & $55.1 \pm 2.1$   & $57.6 \pm 2.0$   & $57.9 \pm 0.9$   & $57.3 \pm 3.3$   \\
 & RF-DETR Nano & 30.5M & 5
   & $14.0 \pm 0.6$   & $22.8 \pm 0.5$   & $23.4 \pm 1.1$   & $22.3 \pm 0.5$
   & $60.7 \pm 0.9$   & $62.3 \pm 0.8$   & $59.3 \pm 1.1$   & $65.6 \pm 1.0$   \\
 & RF-DETR Large & 33.9M & 5
   & $\mathbf{15.4 \pm 1.0}$   & $\mathbf{24.2 \pm 1.0}$   & $\mathbf{24.7 \pm 1.3}$   & $\mathbf{23.7 \pm 1.1}$
   & $61.9 \pm 1.3$   & $\mathbf{63.2 \pm 1.0}$   & $\mathbf{60.9 \pm 0.7}$   & $65.6 \pm 1.4$   \\
 & SAM 3 & 848M & 5
   & $10.4 \pm 0.6$   & $20.6 \pm 0.7$   & $20.5 \pm 0.8$   & $20.7 \pm 0.9$
   & $\mathbf{62.4 \pm 0.7}$   & $62.4 \pm 0.5$   & $58.5 \pm 0.9$   & $\mathbf{66.8 \pm 0.9}$   \\
\hline
\end{tabular}%
}
\vspace{6pt}
\caption{Detection performance of the TFE detector, two YOLO architectures, two RF-DETR variants, and SAM 3, under IoU@50 and IoMin@50. \textbf{Bold} marks the best result per metric and dataset.}
\label{tab:results-combined}
\end{table*}

\begin{figure*}[ht]
\centering

\begin{subfigure}[t]{0.322\textwidth}
    \centering
    \includegraphics[width=\linewidth]{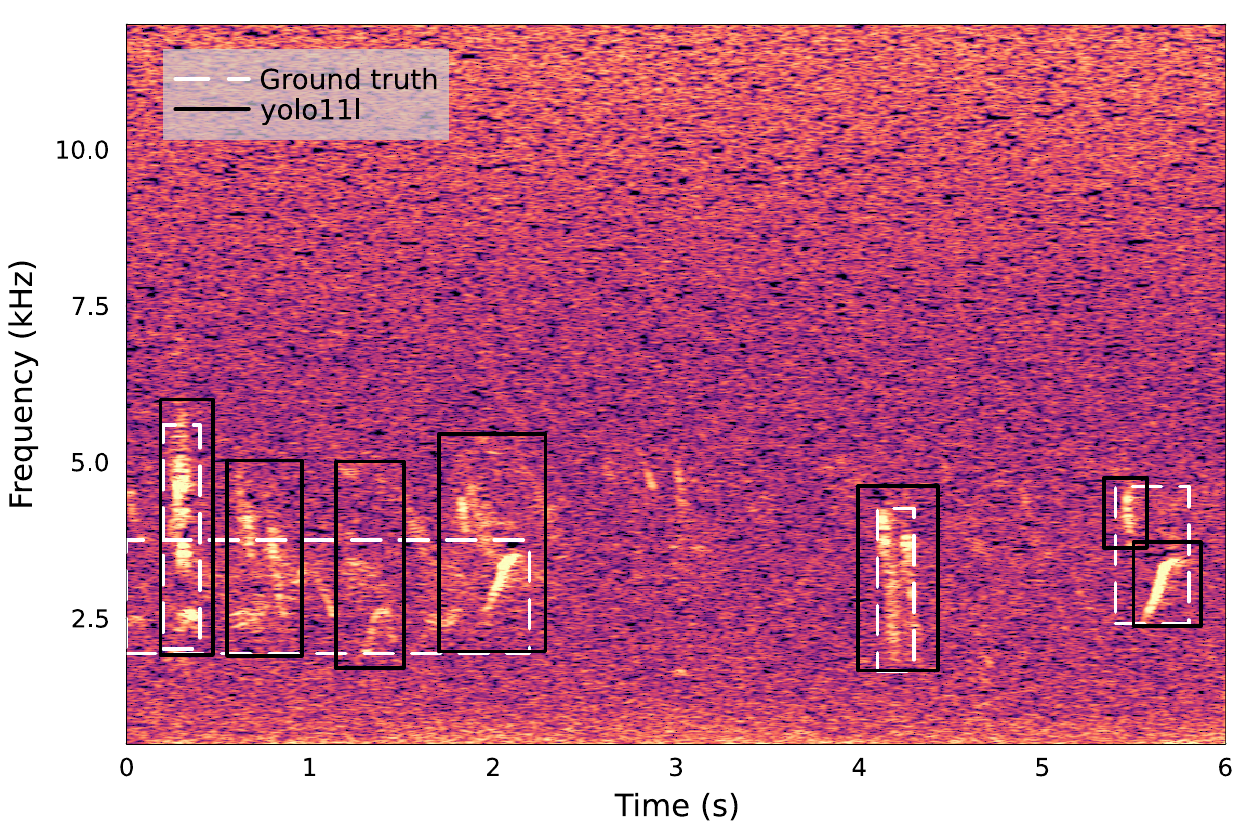}
    \caption{Annotation boundary ambiguity.} 
    \label{fig:yolo-pred-hawaii-1}
\end{subfigure}
\hspace{0.005\textwidth}
\begin{subfigure}[t]{0.322\textwidth}
    \centering
    \includegraphics[width=\linewidth]{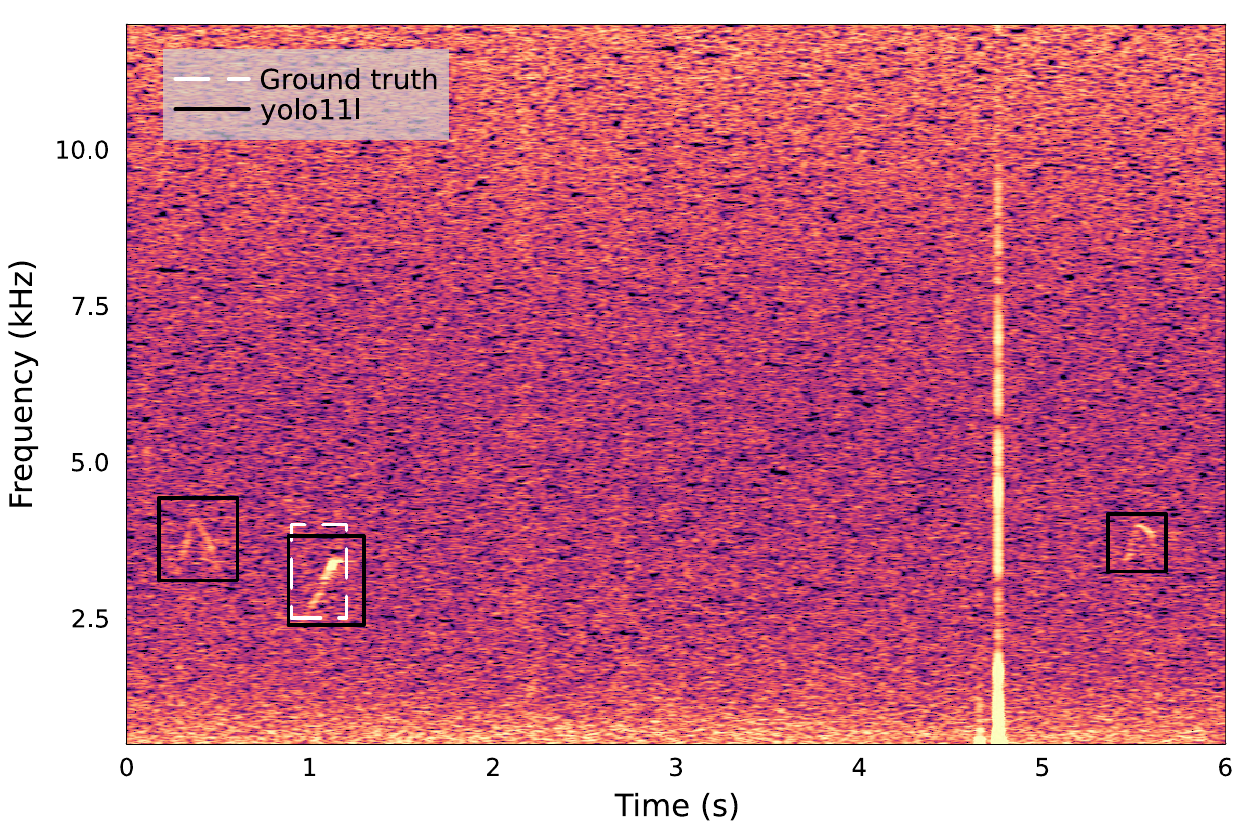}
    \caption{Missing ground-truth annotations.}
    \label{fig:yolo-pred-hawaii-2}
\end{subfigure}
\hspace{0.005\textwidth}
\begin{subfigure}[t]{0.322\textwidth}
    \centering
    \includegraphics[width=\linewidth]{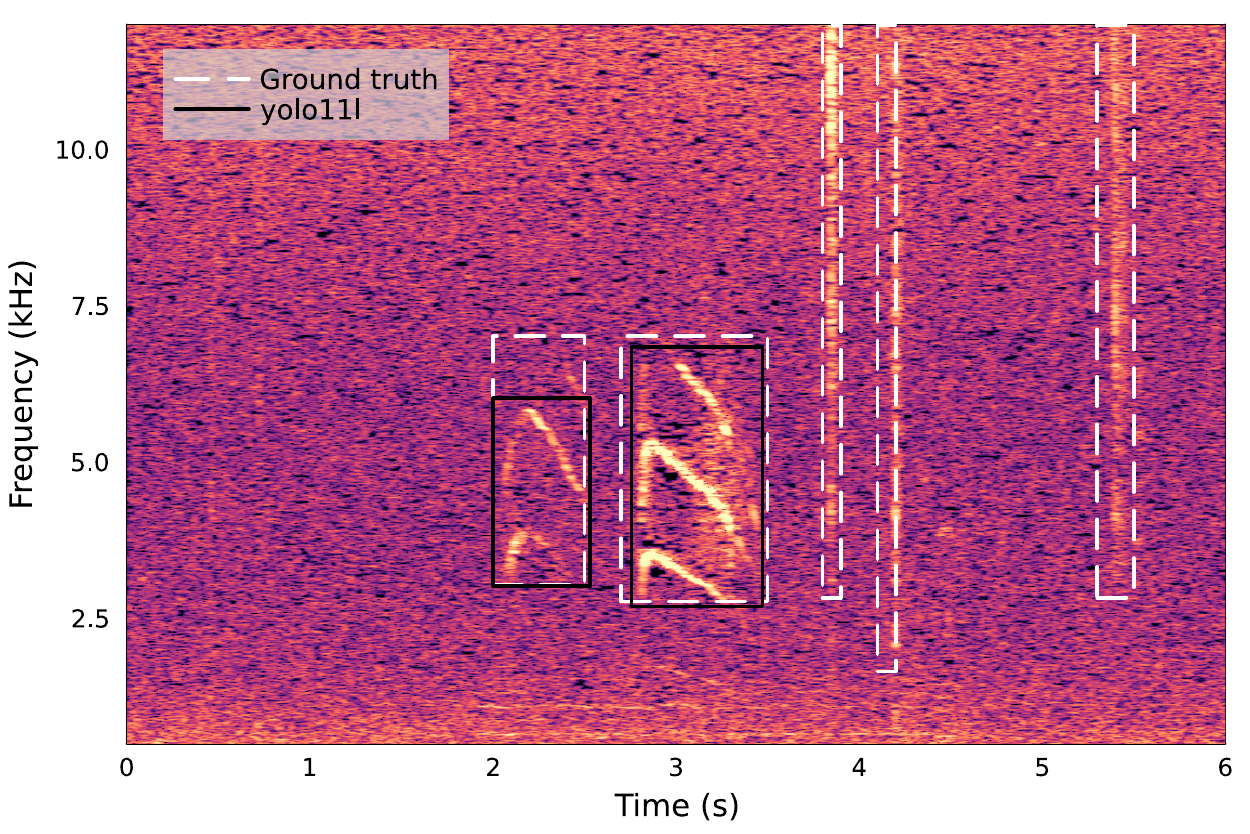}
    \caption{Erroneous ground-truth annotations.}
    \label{fig:yolo-pred-hawaii-3}
\end{subfigure}
\caption{Examples of ambiguous, missing, and erroneous annotations in the Hawaii dataset.}
\label{fig:yolo-pred-hawaii}

\end{figure*}

Performance is reported as mean $\pm$ standard deviation over five training runs (Table~\ref{tab:results-combined}). All three model families outperform the baseline by a large margin across metrics. Figure~\ref{fig:pr-curve} shows the precision-recall curves for the best-performing run of each model, selected by F1-score. Qualitatively, the fine-tuned vision models better suppress non-bird sounds and produce more precise boundaries (Figure~\ref{fig:tfe-fail-sg}). RF-DETR Large performs best across most metrics. On in-distribution data, it yields an IoMin@50 F1-score of 83.4\%, compared with 42.1\% for the baseline. On out-of-distribution recordings, the best model under IoMin@50 depends on the desired precision-recall trade-off: RF-DETR Large performs better at higher recall, whereas SAM 3 performs better at higher precision (Figure~\ref{fig:pr-curve}). We tested fine-tuning SAM 3's vision encoder using Low-Rank Adaptation but it did not improve performance over the frozen encoder.

The models generalize reasonably well to the Hawaii recordings despite differences in species and soundscape from the Singapore training data. However, the highest out-of-distribution IoMin@50 F1 score (63.2\%, RF-DETR Large) remains 20.2 percentage points below its in-distribution score (83.4\%). While part of this gap likely reflects true generalization loss, three annotation-related factors artificially inflate it. First, the datasets differ in annotation practices (Figure~\ref{fig:yolo-pred-hawaii-1}), to which IoMin is less sensitive than IoU. Second, some correctly detected bird calls are absent from the ground truth (Figure~\ref{fig:yolo-pred-hawaii-2}). Third, a few ground-truth annotations correspond to non-bird calls (Figure~\ref{fig:yolo-pred-hawaii-3}).


\section{Conclusion}
In this work, we formulate time-frequency localization of bird calls as an object detection task on spectrograms. Fine-tuned vision models substantially outperform the non-learnable baseline, with RF-DETR Large performing best across most metrics. SAM 3 achieves comparable out-of-distribution performance under IoMin@50. However, its 848M parameters and 3.6~GB disk footprint make it impractical for power- and memory-constrained edge devices typically used in long-duration PAM surveys. In contrast, YOLO11n and RF-DETR Large offer faster inference and require only 5.6~MB and 138~MB, respectively. The results also highlight the need for a standardized benchmark for training and evaluating bioacoustic time-frequency localization methods, which could help separate true generalization loss from annotation-related artifacts such as those observed here. To support research in this area, we release BirdWatch, an open-source analysis and annotation tool for spectrogram-based object detection.

\clearpage
\bibliographystyle{IEEEbib}
\bibliography{ref.bib}

\end{document}